\documentclass[3p,times]{elsarticle}
\usepackage{amsmath,amssymb,amsfonts}
\usepackage{algorithmic}
\usepackage{graphicx}
\usepackage{textcomp}
\usepackage{xcolor}
\usepackage{multirow}
\usepackage{amssymb}
\usepackage{soul}
\usepackage[hyphens]{url}
\def\BibTeX{{\rm B\kern-.05em{\sc i\kern-.025em b}\kern-.08em
    T\kern-.1667em\lower.7ex\hbox{E}\kern-.125emX}}
\usepackage{enumitem} 

\begin{document}

\begin{frontmatter}

\title{RAGOps: Operating and Managing Retrieval-Augmented Generation Pipelines
}

\author[label1,label3]{Xiwei Xu}
\author[label2]{Hans Weytjens}
\author[label1]{Dawen Zhang}
\author[label1,label3]{Qinghua Lu}
\author[label2,label4]{Ingo Weber}
\author[label1,label3]{Liming Zhu}
\address[label1]{CSIRO's Data61, Australia}
\address[label2]{Technical University of Munich, School of CIT, Germany}
\address[label3]{University of New South Wales, School of Computer Science and Engineering, Australia}
\address[label4]{Fraunhofer Society, Munich, Germany}


\begin{abstract}

Recent studies show that 60\% of LLM-based compound systems in enterprise environments leverage some form of retrieval-augmented generation (RAG), which enhances the relevance and accuracy of LLM (or other genAI) outputs by retrieving relevant information from external data sources. LLMOps involves the practices and techniques for managing the lifecycle and operations of LLM compound systems in production environments. It supports enhancing LLM systems through continuous operations and feedback evaluation. RAGOps extends LLMOps by incorporating a strong focus on data management to address the continuous changes in external data sources. This necessitates automated methods for evaluating and testing data operations, enhancing retrieval relevance and generation quality. In this paper, we (1) characterize the generic architecture of RAG applications based on the 4+1 model view for describing software architectures, (2) outline the lifecycle of RAG systems, which integrates the management lifecycles of both the LLM and the data, (3) define the key design considerations of RAGOps across different stages of the RAG lifecycle and quality trade-off analyses, (4) highlight the overarching research challenges around RAGOps, and (5) present two use cases of RAG applications and the corresponding RAGOps considerations. 

\end{abstract}



\end{frontmatter}

\section{Introduction}
\label{intro}

Large Language Models (LLMs) can be instructed through prompting to perform a wide range of tasks, such as programming and translation. A notable trend in their application is the integration of LLMs into compound software systems, which consist of multiple components beyond the core language model~\cite{compound-ai-blog}. Compound LLM systems can perform dynamic behaviors, whereas LLMs by themselves are inherently constrained by their reliance on static training on datasets from some point in time, resulting in fixed parametric knowledge and limited grounding in specific contexts in which the systems are used, such as a given organization. In enterprise settings, 60\% of the LLM compound systems incorporate some form of retrieval-augmented generation~\cite{lewis2020retrieval} (RAG), which improves the relevance, accuracy, and dynamism of LLM outputs by retrieving information from external data. RAG offers a solution to common challenges faced by LLMs, such as hallucinations, outdated data, and the difficulty of removing parametric knowledge, and open up the possibility to access proprietary, internal data from organizations using existing LLMs. By integrating real-time information retrieval, RAG enables continuous updates, possibly incorporating very recent information. RAG systems are compound systems that consist of multiple components blending the LLM’s parametric knowledge with external data retrieval, including, but not limited to \textit{retrieval sources}, \textit{retriever}, and \textit{generator}~\cite{gao2024retrievalaugmentedgenerationlargelanguage}.

LLMOps refers to the practices and techniques used to manage the lifecycle and operation of LLM and LLM compound systems in production environments.
The current state of LLMOps includes a variety of automated tools\footnote{``Managed MLflow,'' Databricks, accessed 5 April 2025, \url{https://www.databricks.com/product/managed-mlflow}}\footnote{``Intelligent Observability,'' New Relic, accessed 5 April 2025, \url{https://newrelic.com/}}\footnote{``LangSmith,'' LangChain, accessed 5 April 2025, \url{https://www.langchain.com/langsmith}} designed to observe, monitor, optimize, and scale LLM applications. The functionality of these tools includes but is not limited to, model versioning, performance monitoring, continuous retraining, inference optimization, and infrastructure scaling to accommodate fluctuating demand. Current LLMOps tools predominantly focus on model and prompting management, offering limited support for data-related aspects, particularly the data created and used after an LLM has been trained and deployed.

This paper introduces RAGOps, a conceptual framework that builds upon and extends LLMOps, with a focus on managing both the data lifecycle and operational aspects of RAG systems. Our contributions are threefold: (1) we conceptualize RAG systems and provide a coherent terminology; (2) we offer design guidelines for building and maintaining RAG systems; and (3) we analyze quality trade-offs to address challenges arising from the continuously evolving data retrieved by RAG applications.
Grounded in the paradigm of \textit{Observability}, RAGOps provides features and functionalities such as managing diverse data structures, formats, and dynamic data updates, while leveraging an array of tools and mechanisms to monitor performance and enable continuous improvement. This is achieved through ongoing evaluation, testing, and the integration of operational feedback. Reliable observability mechanisms are critical, as they directly impact the dependability of LLM applications (including RAG applications). 
Operational aspects are integrated into the system as a whole, and only a comprehensive evaluation and testing of the entire pipeline, including all individual components, can ensure the system's overall reliability.

\begin{figure}
\centerline{\includegraphics[scale=0.2]{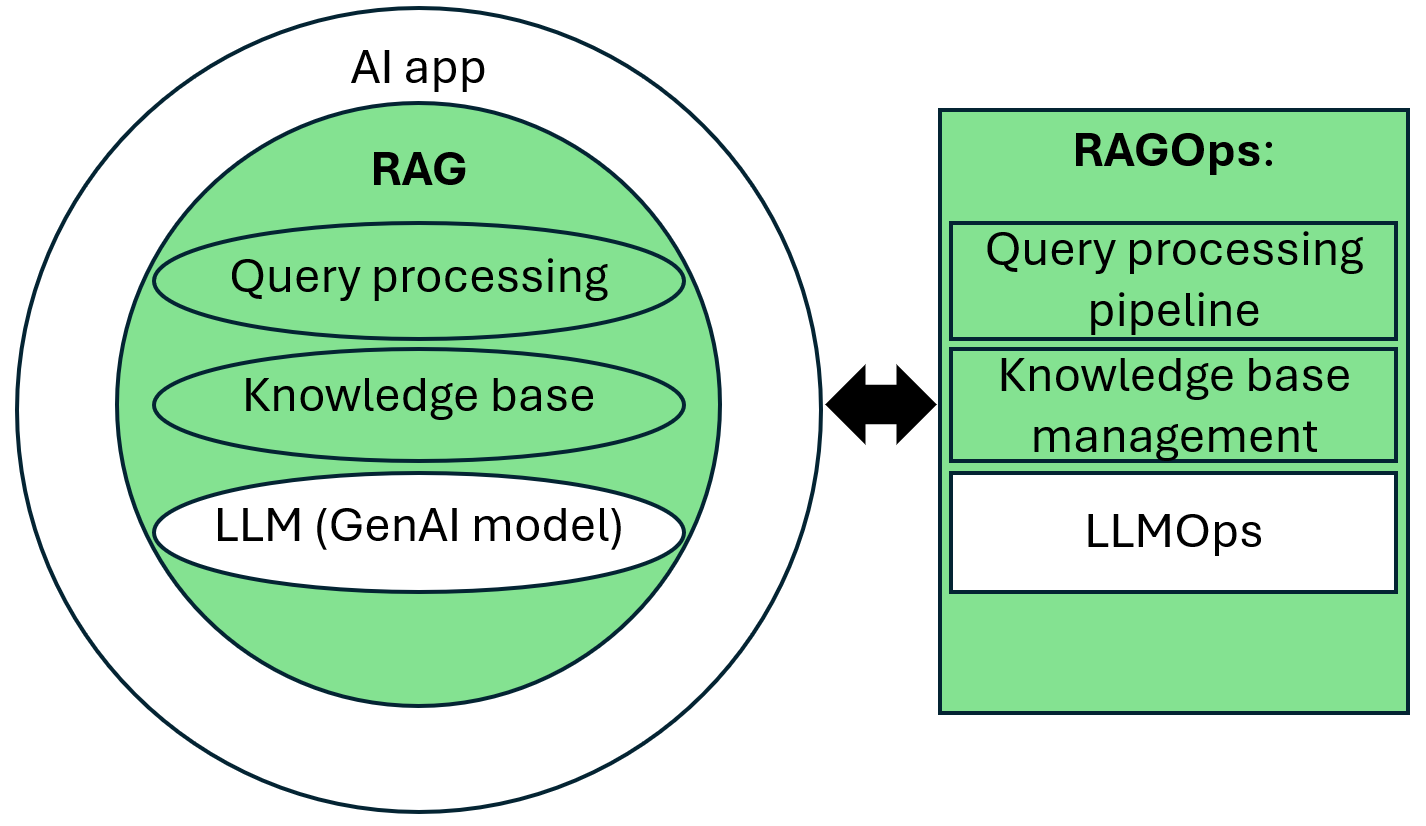}}
\caption{Scoping RAGOps: a RAG system can run either independently or be embedded in an LLM-based compound system (AI App, potentially an AI agent). It has a query processing component interacting with data sources to retrieve information and a genAI model (usually an LLM) to formulate an answer. We exclude LLMOps from our RAGOps discussion.}
\label{fig:positioning}
\end{figure}

Although LLMs \footnote{For simplification, we will refer to \textit{LLMs} in the remainder of this text. For multimodal RAGs, other genAI models will be required.} have their own development pipeline—from data identification and collection to parameterization in an LLM model—this process is outside the scope of this paper. Our focus is on RAG applications that utilize LLMs rather than on training LLMs from scratch. In practice, a RAG system is typically embedded within an AI application such as a chatbot or a customer support assistant, or, more recently an AI agent. However, in the body of this paper we focus on the commonalities in RAG applications and not individual applications, as visualized in Figure~\ref{fig:positioning} -- in contrast, the use cases consider individual applications. The right-hand side depicts our corresponding definition of RAGOps and how this paper is positioned within that framework. Since information retrieval from external data sources is the defining feature of RAG systems, Figure~\ref{fig:positioning} explicitly includes data management. 

In the following section (Section~\ref{related}), we first discuss related work. In Section~\ref{sec_characterization}, we characterize the software architecture of RAG applications that may influence the entailment of RAGOps, based on the 4+1 model view. Next, we outline the lifecycle of RAG architectures, which integrates both the query processing pipeline lifecycle and the data management lifecycle, in Section~\ref{sec_RAGOps}. The data management lifecycle includes stages such as ingest, verify, and update~\cite{davidLifecycle}. In the context of RAG applications, these two lifecycles are closely interrelated, as the LLM and external data sources form the core components of a RAG application. 
We outline the key operational considerations for lifecycle and operational management across the various stages of the RAG lifecycle in Section~\ref{sec_qualities} and address the research challenges in Section~\ref{sec_challenges}, taking into account the distinct characteristics of RAG applications. Additionally, we discuss two use cases in Section~\ref{casestudies}, before summarizing the paper and offering suggestions for future research in Section~\ref{sec_summary}.

\section{Related Work}
\label{related}

\subsection{LLMOps}

The use of LLMs in production environments poses new challenges that expose the limitations of existing methodologies in the DevOps space, like MLOps. LLMOps has emerged~\cite{diaz2024large} with a range of tools and best practices primarily for managing the lifecycle of LLM applications. This includes but is not limited to fine-tuning LLMs, prompt engineering, data provenance for in-context learning, and building infrastructure for training, testing and deploying LLMs. The focus of LLMOps lies in the operational capabilities and infrastructure necessary to evolve LLMS and deploy the new versions effectively~\cite{MLOps4LLM}.

Despite its advancements, most existing LLMOps work~\cite{dong2024agentopsenablingobservabilityllm} primarily targets metrics specifically designed for LLM management and prompt management. Existing work provides limited support for observability in the retrieval process of RAG applications. This gap results in insufficient observability from the perspective of the RAG pipeline.

\subsection{Observability and Monitorability}

\textit{Monitorability} is a quality attribute that refers to the ability to track behavior and performance of software systems using predefined metrics and alerts. It focuses on addressing ``known unknowns'' through actionable insights and alerts~\cite{Charity2018Observability}. While not exclusive to machine learning (ML) and AI systems, monitorability is crucial for their proper operation and sustainment~\cite{lewis2021software}. Automated and continuous improvement processes, including model retraining, rely heavily on effective monitorability.

Monitorability spans the whole lifecycle of AI/ML models, establishing a well-structured pipeline~\cite{STEIDL2023111615} to enhance the performance and other qualities of AI/ML components through iterations of refinement, continuing until the model cannot be further improved. Monitoring changes in external data and understanding their impacts on both individual ML components and the overall AI/ML system introduces additional complexity, highlighting the need for effective and efficient monitoring mechanisms.

\textit{Observability} in AI/ML systems~\cite{shankar2022observability, Charity2018Observability} goes beyond monitoring metrics that are predefined to capture system health. It empowers practitioners to investigate system behaviors by analyzing historical outputs on ``unknown unknowns'' or conducting ``needle-in-a-haystack'' queries~\cite{Gorton2023}. It is essential for LLMOps tools to support observability features, enabling stakeholders to monitor the behavior of the applications, trace the evolution of artifacts, log associated data, detect anomalies, and assign accountability if incidents occur~\cite{lu2024AIEngineering}. 

\section{Characterization of RAG Architecture}
\label{sec_characterization}

Retrieval-Augmented Generation (RAG) is an effective solution for handling hallucinations in LLM by leveraging external data that is not covered by the training data of the LLM. This approach enhances the accuracy and credibility of LLM outputs, especially in knowledge-intensive tasks. Moreover, RAG facilitates continuous data updates and enables the integration of domain-specific expertise into LLM compound systems~\cite{gao2024retrievalaugmentedgenerationlargelanguage, hu2024ragrausurveyretrievalaugmented,wang2024searchingbestpracticesretrievalaugmented}. This section applies the 4+1 View Model of software architecture to conceptualize and characterize the architecture of RAG applications. These architectural characteristics impact the design and implementation of RAGOps (Section~\ref{sec_RAGOps}). 

The \textit{4+1 View Model} of software architecture is introduced by Philippe Kruchten in 1995~\cite{kruntchen1995architectural}. It is a widely used framework in the software engineering and architecture communities. The 4+1 View Model provides a structured approach to describing software architecture through five distinct views, each of which addresses specific concerns from the perspectives of various stakeholders.

\begin{itemize}
    \item \textit{Logic view:} Focuses on the functionality the system provides to end-users.
    \item \textit{Process view:} Addresses the dynamic behavior of the system at runtime, explaining how system processes interact and communicate.
    \item \textit{Development view:} Represents the system from a developer's perspective, emphasizing software organization and management. It is also refereed to as the implementation view.
    \item \textit{Physical view:} Focuses on the deployment of software components within the physical layer and the physical connections between them
    \item \textit{Scenarios:} Uses a set of use cases to describe and illustrate the software architecture. 
\end{itemize}

\begin{figure}[!htbp]
\vspace{5mm}
\centerline{\includegraphics[width=.95\linewidth]{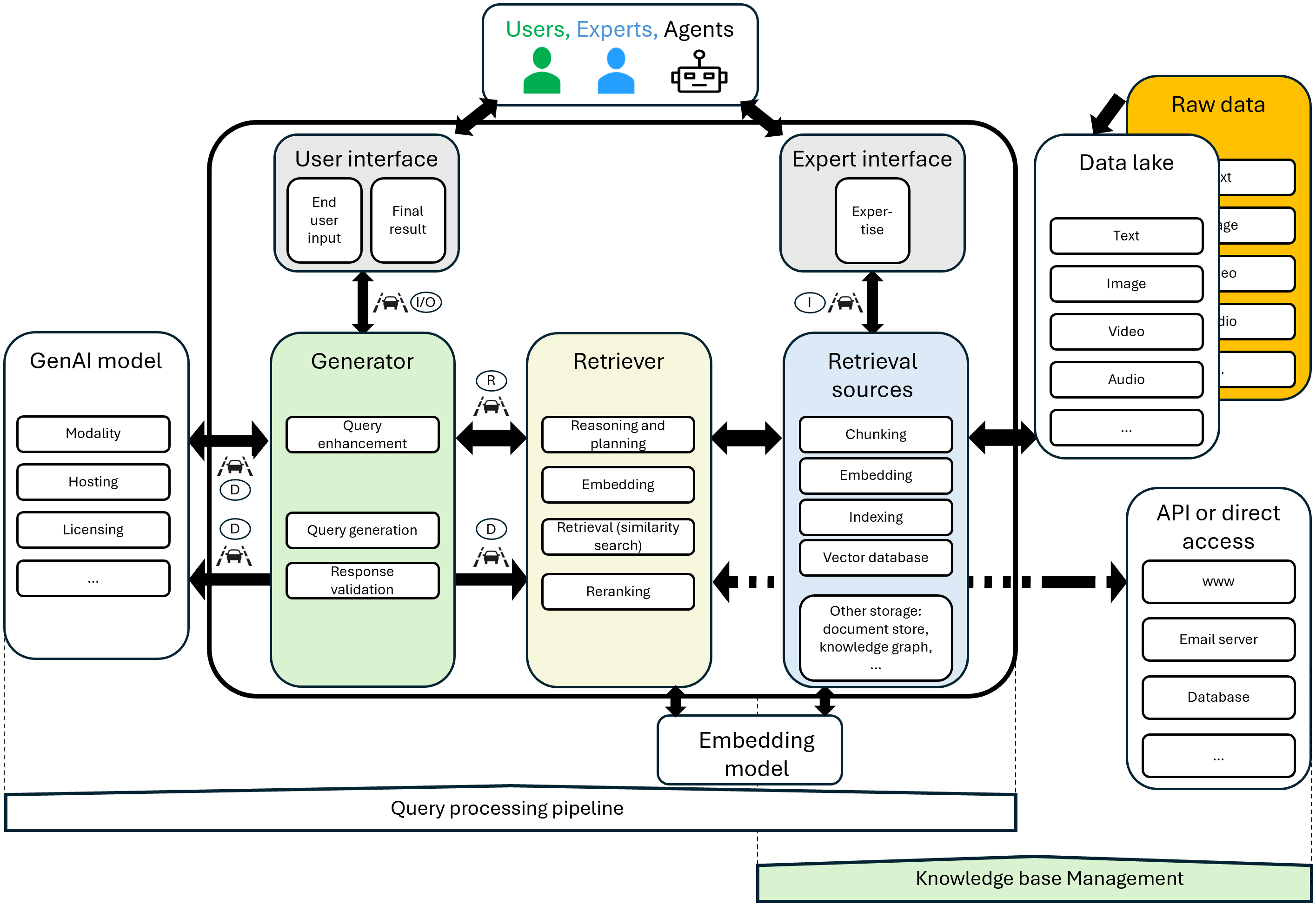}}
\caption{RAG System Architecture.}
\label{fig:rag_architecture}
\end{figure}

A conceptual architecture of a typical RAG system with its key components is demonstrated in Figure~\ref{fig:rag_architecture}.

\subsection{Logic View}

A basic and typical functionality provided by a RAG architecture involves retrieving relevant documents or pieces of information from an external data source to compose the context and then relying the LLM to generate responses with consideration of the retrieved data. As shown in Figure~\ref{fig:rag_architecture}, the three key components of a typical RAG architecture include \textit{retrieval sources}, \textit{retriever}, and \textit{generator}. The user inputs a query through the user interface with the generator potentially optimizing the query. The following retrieval process can be based on embeddings, keyword matching, other search algorithms, or a hybrid approach~\cite{fowler2024genaipatterns}. The retrieval resources may include data sources in different data structures and formats, including but not limited to an embedding vector database and knowledge graph in the retrieval sources module and API access to several internal and external sources. 

Suppose a user (researcher) asks a RAG application about recent developments in quantum computing. The application could first retrieve the latest research papers or articles from an academic database and then generate a summary or explanation based on the retrieved content.

Ensuring the safety and reliability of AI-generated content involves implementing various safeguards \cite{mckinsey2024aiguardrails}. Appropriateness checks screen outputs to eliminate toxic, harmful, or biased content before it reaches users. Accuracy verification confirms that AI-generated information is factually correct and not misleading, combating hallucinations. Regulatory compliance ensures that outputs adhere to relevant laws and industry-specific regulations. User alignment maintains consistency with user expectations and the intended purpose, e.g., ensuring brand consistency. Content validation assesses outputs against predefined criteria.


Human-in-the-loop is a widely adopted strategy in LLM compound systems, where the human is directly involved to intuitively enhance system performance. Human feedback, as a subjective signal, has been used to helping LLM compound systems align with human values and preferences.

In the context of RAG, particularly in specialized domains such as scientific research or law, a portion of knowledge remains undocumented, residing instead in the tacit knowledge of specialists. As shown in Figure~\ref{fig:rag_architecture}, such specialists or experts are one of the user types in the system. For instance, when working with tabular data, the schema may offer a basic structural outline, but the deeper rationale behind the data organization, such as semantic relationships between columns, is not often captured by any document. Subtle characteristics, like patterns of uniqueness, anomalies, or other nuanced data traits, may go unnoticed or unrecorded. Furthermore, contextual insights, such as exponential relationships within the data, are typically uncovered during analysis rather than explicitly provided.

Additionally, analyzing such data effectively often requires specialized expertise. This includes an intuitive grasp of data patterns, the significance of specific variables in relation to carefully designed experiments, and the interplay between variables. Without access to this implicit knowledge and contextual insights, accurate data interpretation can become challenging. In addition to understanding the semantic significance of specific columns in proprietary tabular datasets, other examples of such knowledge gaps include interpreting legal clauses in complex scenarios or under new legislation, and identifying missing connections in graph-based data sources.

In RAG, human feedback can be incorporated as a potential retrieval source, in addition to other passive retrieval sources. The generator can incorporate human feedback into its prompts, providing just-in-time feedback and leading to more informed planning and reasoning. In such cases, when experts (as a type of users) interact with RAG, they become active retrieval sources, accessed through an expert input call.

The RAG system can be used either as a standalone system, as described above, or exposed through an API to an AI agent. Standards to facilitate this new form of communication are currently under development \cite{MCP2025}.

\subsection{Process View}
In addition to the overall architecture, Figure~\ref{fig:rag_architecture} also illustrates the process flow of an advanced RAG-system, outlining the interactions between its key components. It is important to distinguish between two primary processes: the \textit{Query Processing Pipeline} and \textit{Data Management}.

The Query Processing Pipeline refers to the real-time sequence of operations initiated when a user submits a query. Data Management, on the other hand, encompasses the ongoing procedures dedicated to organizing and updating the system's data and retrieval sources.

\subsubsection{Query Processing Pipeline}
In advanced RAG-systems, the generator may contain a module that enhances the user's query to optimize retrieval. The retriever's reasoning and planning module analyzes the (enhanced) query to determine the most effective retrieval strategy. This decision-making process includes selecting appropriate retrieval sources (e.g., vector database within the retrieval sources module, external APIs, direct access to a proprietary database, or consulting the human expert), picking specific retrieval methods (e.g., similarity search, keyword search), and formulating queries for various APIs (e.g., general search engines, specialized databases). 

After its decision-making process, the retriever performs the actual retrieval. In the regular cases, it consults the vector database. That requires a preliminary embedding of the (enhanced) user query. The reranking module prioritizes the retrieved information, removing redundancies and irrelevant data points. In a more agentic approach, the autonomous retriever could also decide to reiterate and refine the whole process after detecting insufficient results. Both reasoning and planning and reranking can involve heuristic rules, ML models or interaction with an LLM (as indicated by the arrow connecting the retriever and the LLM in Figure~\ref{fig:rag_architecture}). 

The generator synthesizes a query integrating the original user query with the retrieved information, which is then sent to a selected LLM. The model's response is validated to verify accuracy and relevance before sending it back to the user interface. When the system has agentic capabilities, it can also decide to reiterate and improve the process, either in its entirety or for a selected step.

The interaction between the retriever and generator can also become more intertwined, a process referred to as Retrieval Interleaved Generation (RIG)~\cite{RIG}. In RIG, real-time information retrieval is dynamically embedded within the LLM's generation process. Unlike the linear approach, where external data is retrieved before the LLM generates a response, RIG allows for seamless integration of retrieval into generation, enabling finer-grained interactions between the two components. 

Early RAG systems employ a static retrieval process, where the retriever and generator are used linearly. Recently, RAG systems have progressively evolved toward more dynamic workflows, incorporating adaptive, recursive, and interactive retrieval processes. These advancements are supported by emerging RAG frameworks~\cite{zhao2024RAG, gao2024retrievalaugmentedgenerationlargelanguage}.

\subsubsection{Data management}

The retrieval sources module processes data differently depending on the modality: texts are chunked to split long documents into smaller, semantically meaningful units before being converted into embeddings, while other modalities such as images, audio, and video are directly embedded without chunking. Embeddings are stored in a vector database for similarity-based retrieval, while other storage solutions (e.g., document stores, knowledge graphs) can be used for alternative retrieval methods such as keyword-based or structured queries. The mechanisms of data management which keeps the data lake and retrieval sources current will be discussed in detail in Section~\ref{subsec:knowledgebase}.

\subsubsection{Guardrails}
Within the RAG-context, guardrails can be implemented at different stages of the pipeline \cite{shamsujjoha2024taxonomy, lu2024agentarceval, ganju2024securemedicalapps, smith2025securing} as illustrated by the \includegraphics[width=1.5em]{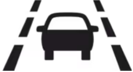} icons in Figure~\ref{fig:rag_architecture}. Guardrails serve as intermediaries between two components when they interact.

Input rails (I) process user inputs, either rejecting them to halt further processing or modifying them, such as masking sensitive information or rephrasing before handing them over to the generator. Dialog rails (D) shape the prompting of the LLM and guide the conversation's progression. Retrieval rails (R) manage the retrieved data segments. They can exclude retrieved information from being passed to the retriever or from being used by the generator to prompt the LLM or modify segments, for instance, to conceal sensitive information or poisoned information. Output rails (O) apply to the LLM's generated outputs, with the capability to reject responses, preventing them from reaching the user, or to modify them, such as by removing sensitive content. More rail actions and targets can be found in \cite{shamsujjoha2024taxonomy}, but it is important to balance the number and placement of these guardrails to avoid redundancy, which can lead to unnecessary complexity and inefficiency.

\subsection{Development View}

Embeddings in vector databases are the most widely used and typical retrieval sources for RAG. 
Beyond embeddings, external databases and other data sources can be structured and accessed in various ways. Humans with tacit knowledge can also serve as 
more active retrieval sources. 
Below is a list of possible retrieval architectures from a developer's perspective, which shows the flexibility of LLMs in integrating with external data sources to deliver accurate, context-aware, and real-time responses. Each architecture can be refined 
to specific use cases depending on the characteristics of the required information and the available technical infrastructure.

\begin{itemize}
    \item \textbf{RAG-system retrieval sources} contain data from the curated data lake that are processed and stored in a RAG-specific way. Commonly, the information is stored along embeddings in a vector database. Alternatives include knowledge graphs (text or images may be associated with the nodes of the graph) and document stores (to retrieve text chunks via complementary retrieval mechanisms such as keyword matching TF/IDF), etc.
    \item \textbf{APIs} permit direct access to (possibly non-curated) data to both internal sources, such as databases, email servers, etc., and external sources, such as the web.
    \item \textbf{Human-in-the-retrieval} RAG applications integrate humans' knowledge in several ways. One approach is for the human experts to manually provide additional context through prompting. Alternatively, RAG applications can connect to external crowdsourcing knowledge platforms via APIs to integrate collective expertise.
\end{itemize}

\subsection{Physical View}

There are various deployment and infrastructure options for RAG systems. Each key component (generator, retriever, and retrieval sources) requires specific deployment and hosting considerations, such as whether to host the component on the cloud or on-premises.

Selecting the appropriate LLM for RAG systems is akin to choosing computational infrastructure, particularly in scenarios involving multiple LLMs for different purposes. For instance, one LLM may be used for general generation while another, such as a dedicated ``LLM guard,'' verifies content at various stages. An example is reviewing the execution of SQL queries that access external tabular data~\cite{pedro2023promptinjectionssqlinjection}. 

Retrieval sources may need to be constructed from raw data in various formats as shown in Figure~\ref{fig:rag_architecture}. Alternatively, existing data sources can be utilized directly through RESTful APIs, such as knowledge graphs like Wikidata\footnote{Wikidata, accessed 5 April 2025, \url{https://www.wikidata.org/}}.

\subsection{Scenario}

The usage scenarios of RAG systems vary widely, falling primarily into two categories or a combination of the two: 1) Querying specific details, and 2)
Gaining a holistic understanding of a source~\cite{edge2024localglobalgraphrag}. The source in question could be in different format, a text-based document, an image, or a video or a combination of them. An example of querying specific details in the context of a customer assistant might be, \textit{``What is the return policy for online purchases?''} Similarly, in the context of a legal assistant, a user might ask, \textit{``What's the capital gains tax on the sale of a rental property?''} Tasks that require a holistic understanding focus on broader comprehension or synthesis of information from a big document or multiple sources. For example, a legal assistant might be tasked with summarizing a lengthy legal document to extract key clauses, implications, and risks.

Some usages scenarios require a combination of both querying specific details and gaining a holistic understanding, particularly in reasoning-intensive tasks. For instance, in a scientific research assistant scenario, a researcher might query specific data points from different studies while simultaneously synthesizing these details to form a cohesive understanding of a broader research landscape. 

In such reasoning tasks, the RAG application must seamlessly integrate detailed queries with broader context comprehension, ensuring accuracy and relevance at every stage. Depending on the characteristics of the usage scenarios, different data sources with various data structures are selected and/or structured.

\section{RAGOps}
\label{sec_RAGOps}

Excluding LLMOps and the wrapping app (see Figure~\ref{fig:positioning}), RAGOps can be broken down into two tightly knit, yet different lifecycles: the query processing pipeline, that follows the traditional DevOps lifecycle and a more data-centric data management. 

\subsection{Query Processing Pipeline}
When stripped of its surrounding application, ignoring the LLM(s) that drive it, and setting aside the separate concerns of data management, a RAG system becomes a typical piece of software. As such, it naturally falls within the domain of traditional DevOps practices.

\begin{figure}[ht]
    \centerline{\includegraphics[width=\linewidth]{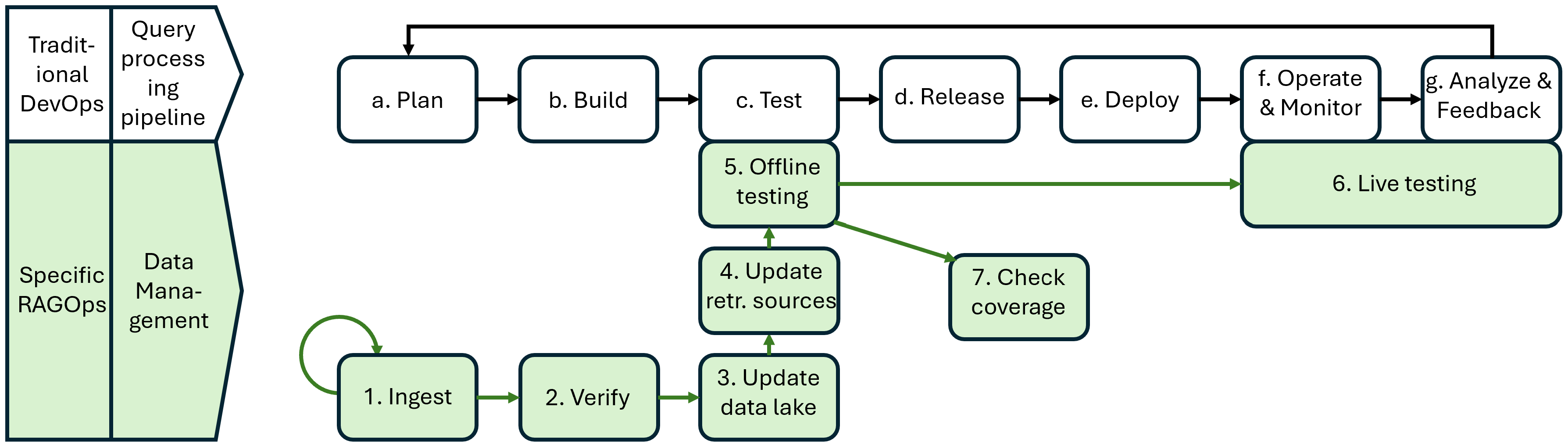}}
    \caption{RagOps: intertwined query processing pipeline and data management lifecycles.}
    \label{fig:devops}
\end{figure}

The DevOps lifecycle is comprised of seven interconnected phases in a continuous loop, as illustrated in the upper (white) part of Figure~\ref{fig:devops}. It embraces the principles of continuous integration and continuous deployment.

\begin{enumerate}[label=\alph*.]
    \item \textbf{Planning and Development}: This phase involves defining system requirements, selecting appropriate tools, and outlining an architectural blueprint. For RAG systems, this means identifying the knowledge domain, choosing data sources, and designing retrieval and generation strategies. Teams determine the LLMs, embedding models, vector database and other retrieval sources, prompt engineering techniques, etc.

    \item \textbf{Build}: The build phase includes setting up necessary components, writing core logic, and ensuring dependencies are properly configured. In a RAG system, this involves preprocessing and chunking documents, generating embeddings, indexing them in a vector database for efficient retrieval, and constructing  knowledge graph in the case of using knowledge graph. Additionally, teams configure the retrieval pipeline, integrate it with the LLM, and package components—often using containerization.

    \item \textbf{Testing}: While standard testing covers functionality, performance, and security, RAG-specific evaluation also includes assessing retrieval relevance, response accuracy, and overall coherence. This means verifying that embeddings align with expected semantic meanings, ensuring retrieved documents are the most relevant, and evaluating the model’s factual consistency. System load testing is particularly crucial, as retrieval latencies and model response times can impact real-time performance. Benchmark datasets and automated testing frameworks are commonly used to validate pipeline effectiveness before deployment.

    \item \textbf{Release}: The release phase finalizes the system for production, ensuring all components are properly integrated and meet performance, accuracy, and stability requirements. This involves validating the compatibility of the retriever, generator, and 
    retrieval sources, refining versioning strategies, and preparing documentation. The system undergoes final approval, often deployed in a staging environment before full deployment, ensuring updates can be rolled out smoothly.

    \item \textbf{Deployment}: Deployment focuses on infrastructure stability, automated scaling, and continuous integration. In RAG systems, deployment also involves ensuring seamless interaction between the retriever, 
    retrieval sources, and language model in real-world conditions. The ongoing challenges concerning data updates could be included in the deployment phase as well. Given their frequency and importance, we argue for a treatment as a separate component of the RAGOps lifecycle.

    \item \textbf{Operate and Monitor}: Monitoring covers infrastructure health, logging, and alerting. In a RAG system, additional tracking is necessary for retrieval efficiency, response fidelity, and data drift detection. Monitoring tools assess whether retrieved documents remain relevant, whether response quality degrades due to evolving data sources, and how changes in embeddings impact retrieval precision. The system is continuously adjusted to maintain effectiveness as the underlying data evolves.

    \item \textbf{Analyze and Feedback}: Incident response and performance tuning are core DevOps practices, but in RAG, this phase also involves diagnosing retrieval errors, detecting data drift, and refining query-document matching. Teams analyze retrieval logs, user interactions, and model responses to identify systemic issues, such as retrieval failures, hallucinations, or outdated information. Insights from this analysis guide iterative improvements, ensuring that the system remains accurate and effective over time.
\end{enumerate}

Once the query processing pipeline operates in the production environment, analysis based on measurements taken will lead to occasional updates, effectively closing the loop and triggering new developments. We elaborate on the measurements in our discussion about monitorability and observability in Section \ref{subsec:RAG_pipeline}. However, during deployment, the data will be updated at a much higher frequency, potentially continuously. Data maintenance requires a dedicated lifecycle.

\subsection{Data management}\label{subsec:knowledgebase}
As a highly dynamic core component of the RAG system, its data demands careful management. Data entering it needs to be checked to upkeep the system's performance. Shifts in the data distribution and/or system usage patterns might also impact the system's end-to-end performance, requiring evaluations of the RAG system's responses. These evaluations call for a test data set, whose coverage needs checking as well.

\subsubsection{Ingestion}
Data can be ingested from diverse sources such as file systems, databases, APIs to web services, the web, real-time feeds, human experts, etc. Data can be structured or unstructured. It exists in text-based formats (e.g., JSON, Markdown), in document formats (e.g., PDF, .doc), in image formats (e.g., JPEG, GIF), as structured data (e.g., relational SQL-databases, knowledge graphs, graph databases), as audio (e.g., MP3) and video (e.g., MOV) files, etc.

Depending on the nature of the data, methods to automatically pull in the data can be chosen from webhook triggers, change data capture (CDC), polling mechanisms, file system watchers, stream processing, web crawlers, etc. Manual inputs (e.g., continuous learning) are equally possible. Most of these methods can also be implemented in push rather than pull mode. The frequency of these data transfers ranges from near-continuous to infrequent. Note that, in our context, ingestion also includes deletions and updates of data.

Additionally, ingestion includes necessary format conversions, such as extracting text from PDFs, generating descriptions for images, or transcribing audio, ensuring that data is ready for verification and later processing.

\subsubsection{Verification}\label{subsubsec:verification}
Once the data is ingested, it must be systematically verified to remain accurate and consistent. This verification process contains multiple checks:

\begin{itemize}
    \item \textbf{Quality} The data should be well-structured and not contain corrupted or malformed content. Where relevant, the correctness of the data can be verified through external validation.
    \item \textbf{Completeness} All expected fields, including metadata (e.g., timestamps, authorship, source), should be present and contain values. Texts should not be truncated.
    \item \textbf{Recency} Older versions should not be retained, but archived and replaced by more recent versions. Data recency is evaluated using timestamps from the metadata, embedded timestamps, or semantic analysis.
    \item \textbf{Consistency} Contradictory information should be detected, e.g., by using semantic similarity checks (embeddings, cosine similarity). To resolve conflicts, the data sources can be weighted for trustworthiness and/or human intervention can be solicited.
    \item \textbf{Uniqueness} Duplicates should be eliminated, for example with hash-based deduplication. To discover semantic redundancies, the system can calculate embeddings followed by clustering. Term frequency-inverse document frequency (TF-IDF) and Latent Dirichlet Allocation (LDA) can be used to identify documents discussing similar topics.
\end{itemize}

External sources to which the retriever has direct API access pose a special challenge. In most cases, the checks described above would lead to unacceptable latencies, as these external sources are consulted in real time. Retrieval guardrails, however, can help to enhance veracity.

\subsubsection{Updating data lake}
This phase is about storing the previously ingested and verified data efficiently in a data lake. In organizations where multiple RAG systems or applications wrapped around RAG systems share overlapping data sources, maintaining a separate data lake (see Figure~\ref{fig:rag_architecture}) before chunking and embedding in the ``retrieval sources'' component offers several advantages. By centralizing the ingested and verified data, the data lake ensures consistency, reduces redundancy, and provides a single source of truth across systems. This architecture enables different RAG systems, or other downstream applications, to access pre-validated data without duplicating ingestion and verification efforts. Additionally, it supports better data governance, facilitates auditing, and allows for easier updates across multiple systems. 

In this stage, the updates and deletions identified in the previous verification phase are now applied to the data lake. This includes removing duplicates, replacing outdated records, and correcting errors. Instead of outright deletion, archiving old versions allows for rollback if needed and helps maintain a historical record of changes. Each modification must be versioned to maintain a clear history of changes, to allow the system to revert to earlier versions if necessary. When updates are applied, the data must remain logically sound (e.g., free of orphans). This means that when new or updated data is added, any related information should align correctly, and previous references to outdated data should be adjusted accordingly. Additionally, access control metadata should be stored alongside each data entry, ensuring retrieval queries respect access permissions and enforce role-based restrictions where necessary. 

\subsubsection{Updating retrieval sources}
Next, the data changes are propagated to the retrieval sources block in the pipeline, where incremental chunking and embedding are performed on newly added or modified data. Indexing is also updated accordingly, ensuring that deletions and outdated information are properly removed from the retrieval process.

\subsubsection{Offline testing}
\label{sec:offline_testing}
The offline testing of the RAG system can be triggered by RAG-pipeline changes or data distribution shifts. Further to the initial setup, RAG pipeline changes concern modifications of the pipeline components, e.g., switching to another LLM (version), adapting the query generator, adopting another chunking method, or swapping the embedding algorithm. Data distribution shifts can happen when new data sources from other domains are attached to the system, e.g. an API to a new database, an additional set of URLs to the web crawler, or the inclusion of another language. Gradual, latent data distribution shift will happen over time as well and should be monitored. The tracing of these changes relates to and supports the quality attributes laid out in Section~\ref{subsec:RAG_pipeline}, especially monitorability and observability.

We propose a multi-layered platform for offline testing (as well as live testing, see Section~\ref{subsubsec:liveltesting}) which happens at three levels of granularity: the module (e.g., chunking),  the component (e.g., retrieval), and the system level (end-to-end).  After unit testing a module, its effect on the component, and end-to-end system are tested as well. Each level of testing granularity requires different test data and metrics. Metric selection should align with end-user priorities. Table~\ref{tab:rag_testing} illustrates the testing platform with the example of embedding. It differentiates between traditional DevOps testing and specific RAG-testing. For a detailed discussion on metrics, refer to existing work \cite{chang2024retrieval}.

\begin{table}[ht!]
    \centering
        \caption{Example of usage of the three-level RAG testing platform. Module-level embedding affects both the ``retrieval'' and ``retrieval sources'' components, so these should be tested together at the second, component level before testing the end-to-end system.}
    \renewcommand{\arraystretch}{1.3} 
    \setlength{\tabcolsep}{4pt} 
    \begin{tabular}{|p{2.5cm}||p{3cm}|p{3cm}|p{3.2cm}|p{3.2cm}|}
        \hline
        \textbf{Level} & \textbf{What We Test} & \textbf{Test Data} & \textbf{Traditional DevOps} & \textbf{RAG-Specific} \\
        &  &  & \textbf{Metrics} & \textbf{Metrics} \\
        \hline\hline
        \textbf{Module}: Embedding quality & Similarity, drift, efficiency & Benchmark datasets, synthetic tests & Embedding generation time (ms), memory usage per embedding & Cosine similarity, embedding drift, t-SNE/UMAP Clustering \\
        \hline
        \textbf{Component}: Retrieval quality & Rank relevance, resilience to noise & Public RAG retrieval evaluation benchmarks (BEIR, MS MARCO), production queries & retrieval latency (ms), Index update speed & Mean reciprocal rank (MRR), recall@K, precision@K, normalized discounted cumulative gain (nDCG) \\
        \hline
        \textbf{End-to-End} Response quality & Faithfulness, hallucination, correctness & Human-labeled, production logs & Full pipeline latency, response generation time & Faithfulness score, hallucination rate, BLEU, ROUGE, BERTScore\\
        \hline
    \end{tabular}

    \label{tab:rag_testing}
\end{table}

The RAG-specific, data-oriented testing calls for domain-specific test datasets. The relevance of these test data sets must be monitored against actual system usage (see\textit{~\ref{coverage}. Coverage checking}). When introducing new data domains, languages, or formats, the test sets should also be updated or expanded beforehand to reflect these changes. Newly added test sets ensure the RAG system performs effectively on the new data, while existing test sets remain crucial to verify that performance on previously covered domains does not degrade. This dual approach helps maintain both forward adaptability and backward compatibility.

In addition to these domain-specific test sets, standard benchmarks can also play a valuable role. Widely-used retrieval and generation benchmarks (e.g., BEIR, MS MARCO) provide a consistent baseline for evaluating pipeline or data changes, enabling comparisons beyond the  organization’s domain-specific test sets. Synthetic data can also complement testing by enabling controlled simulations of edge cases, rare scenarios, or diverse query variations that may be underrepresented in the test sets or benchmarks.

RAG systems introduce significant stochasticity in both retrieval (e.g., approximate nearest neighbor search or ranking tie-breaking) and generation (LLM outputs). This variability makes testing inconsistent, as the same query may yield different retrieved documents or generated responses across runs, complicating reproducibility and metric comparison. To mitigate this, fixing random seeds will ensure more consistent and comparable test results.


The offline testing phase may result in the exposure of the need for modifications to the RAG system, such as swapping the LLM, updating prompts, embeddings, etc. Such modifications, in turn, will trigger further testing before being released and deployed.

\subsubsection{Live testing}
\label{subsubsec:liveltesting}
After a positive evaluation in the offline testing phase, the RAG system is released and deployed. To minimize risk, the following deployment strategies can be considered:
\begin{itemize}
    \item \textbf{Shadow testing} Running a new version of the pipeline in parallel with the live version without affecting users, to compare outcomes before switching; this approach allows testing both individual components (e.g., retriever or generator) and the complete pipeline.
    \item \textbf{A/B testing} Deploying the new system to a small, but representative subset of users while keeping the old pipeline for others.
    \item \textbf{Staged roll-out} Deploying the new pipeline incrementally to user subsets.
\end{itemize}

The deployment marks the start of live testing to achieve the quality attributes (see Section~\ref{subsec:RAG_pipeline}). The objective of live testing is to validate the system’s functionality, performance, and user experience in a real-world environment, identifying issues that may not surface during controlled offline testing.

From a data-centric perspective, live testing is a continuous exercise, as gradual changes to the RAG data occur over time, and usage patterns—such as the types of queries—may shift. Continuous monitoring and iterative adjustments are therefore essential to maintain system reliability and relevance. Similar to offline testing, live testing can be conducted at three levels: module-level (testing individual modules, such as query generation or reranking, in isolation), component-level (evaluating integrated components or sub-parts of the pipeline, such as generation or retrieval), and end-to-end-level (assessing the entire RAG system under real-world conditions).

Metrics play a central role in live testing to monitor system performance, user experience, and data consistency. Next to the ``traditional'' metrics such as latency and throughput (to ensure responsiveness under real-world load), RAG-specific metrics include retrieval quality (e.g., precision, recall, and relevance of retrieved documents), generation quality (e.g., response accuracy, coherence, and hallucination rate), and usage analytics (e.g., query distribution shifts and user engagement). Continuous tracking of these metrics enables timely detection of data drift, performance degradation, and evolving user needs.

To ensure timely detection of performance degradation, predefined thresholds should be established for key metrics during live testing. When one or more metrics exceed these thresholds, it signals the need for further investigation and optimization of the RAG system. This triggers a feedback loop, where the RAG system is revised and then sent back to the offline testing phase before proceeding to the life testing phase again.

\subsubsection{Coverage checking}\label{coverage}
Coverage checking ensures that the test data used during offline testing remains representative of the live usage patterns observed during live testing. Over time, several factors can cause the test data to become outdated or misaligned with actual user interactions. Changes in user behavior, shifts in query distributions, the addition of new data domains, languages, or content sources, and the introduction of new system components (e.g., a different chunking or embedding method) can all lead to discrepancies. Without adequate coverage checking, the RAG system may perform well on outdated test sets but fail to meet real-world user needs, risking degraded retrieval accuracy, hallucination issues, or slower response times.

Coverage checking can be performed at multiple stages of the RAG system to capture different aspects of representativeness:

\begin{itemize}
    \item \textbf{Query Coverage}
    At the query generation and retrieval stages, comparing live user queries with test set queries is essential. Techniques such as TF-IDF vector similarity or embedding-based comparisons (e.g., cosine similarity using query embeddings) can quantify how closely the test set aligns with current usage. Low similarity scores or newly emerging query clusters (detected via dimensionality reduction methods like t-SNE or UMAP) indicate gaps in coverage that require test set updates.

    \item \textbf{Retrieval Coverage}
    In the retrieval component, coverage checking assesses whether the retrieved documents for live queries reflect those considered during offline testing. If live queries consistently retrieve documents outside the scope of the test set, this signals that the data has evolved or user needs have shifted. Using retrieval metrics like recall@K on both test and live queries can help track this drift.

    \item \textbf{Generation Coverage}
    At the generation stage, coverage checking examines whether generated responses from live queries align with the topics, intents, and styles covered in the test set. Variations can be detected through semantic similarity analysis using embedding models or by applying topic modeling (e.g., LDA) to compare topic distributions.

    \item \textbf{Vocabulary and Domain Coverage}
    Tools like TF-IDF are particularly useful for identifying vocabulary drift. By comparing term frequencies between test and live data, new or underrepresented terms can be flagged for inclusion in updated test sets. This approach is beneficial when new data sources or languages are integrated into the RAG system.
\end{itemize}

In practice, coverage checking should be continuous and automated where possible, with predefined thresholds triggering reviews when significant discrepancies arise. For example, if less than 85\% of live queries achieve a certain similarity score to test queries, the system should prompt a test set expansion. By maintaining strong coverage alignment, the RAG system ensures both forward adaptability (supporting new user demands) and backward compatibility (retaining performance on existing use cases).

\section{Impact of RAGOps on Quality Attributes}
\label{sec_qualities}

RAGOps encompasses the techniques and mechanisms for managing, monitoring, and optimizing interactions between LLMs and external data sources and databases, addressing both retrieval and generation processes as well as the characteristics of the retrieval sources. The design and implementation of RAGOps has impact to the qualities of RAG systems. This section examines the impact of RAGOps on various qualities, including the factors that influence these qualities.

\subsection{RAG system Qualities}\label{subsec:RAG_pipeline}

RAGOps aims to establish a CI/CD infrastructure that supports the full lifecycle of RAG, including both the query processing pipeline and data management as outlined in Section~\ref{sec_RAGOps}. This infrastructure automates the RAG system, facilitates seamless connections, and implements feedback loops to enhance performance of RAG applications, including improved observability and monitorability. Table~\ref{tab:qualities} highlights the quality attributes of a RAG system that must be considered, along with the corresponding operational design considerations.

\begin{table}[!htbp]
\caption{RAG system Qualities}
\label{tab:qualities}
\begin{center}
\begin{tabular}{|c|l|}
\hline
\textbf{\textit{Quality}} & 
\textbf{\textit{Operational Design Consideration}}\\
\hline
\hline
\multirow{2}{*}{Adaptability} &   Data structure construction \\
& Granularity configuration \\
\hline
\multirow{3}{*}{Monitorability} &  Performance degradation detection: Accuracy, Latency, Relevance\\
& Data source change detection \\
& Clear, consistent, and meaningful performance metrics \\
\hline
\multirow{2}{*}{Observability} & Retrieval: Injection attack detection, Relevance\\
& Generation: Context adherence, Toxicity,Usefulness \\
\hline
 \multirow{2}{*}{Traceability} & Versioning\\
 & Retrieval routing\\
 & Grounded source\\
\hline
\multirow{2}{*}{Reliability}  & Resource usage\\
 & Bandwidth\\
\hline
\end{tabular}
\end{center}
\end{table}

\textbf{Adaptability.} Adaptability is a critical quality for managing the rapid technical evolution of the RAG system. As discussed earlier in Section~\ref{sec:offline_testing}, changes to the RAG system involve modifications to all its components, such as switching to a different LLM, adjusting the query generator, adopting alternative chunking methods, or replacing the embedding algorithm. To ensure adaptability, the RAG pipeline is designed to support automated updates to retrieval sources whenever raw data changes or performance degradation is detected. Development design choices, such as the method of knowledge graph construction and the granularity of retrieval units, influence the efficiency of adaptation.

\textbf{Monitorability.} Monitorability is essential for maintaining the performance of retrieval sources. It requires components capable of detecting performance degradation and changes in the data sources used by these retrieval sources. Clear, consistent, and meaningful metrics must be defined to assess performance and utility. 

\textbf{Observability.} RAG systems require comprehensive observability, including their operational functionality, to ensure reliability.
An observability infrastructure layer is essential and must span across all components. This layer logs all queries, responses, and the inputs and outputs of each component, providing comprehensive visibility into the application's operations and enhancing monitoring and evaluation capabilities.

The retriever's functionality primarily relies on a series of query tools designed to retrieve information from external sources in various data structures and formats without modifying their states. Beyond performance metrics such as latency and relevance, observability must also detect potential injection attacks that could potentially compromise retrieval sources. For instance, when retrieving tabular data, the SQL queries generated by the LLM should be thoroughly validated to prevent SQL injection.

After retrieval, the generation functionality adjusts the context retrieved from external sources before sending it to the LLM model to answer users. Similarly to the retriever, the generator needs to monitor metrics in addition to performance, like toxicity and context adherence.

\textbf{Traceability.} From a data-centric perspective, traceability is paramount in a dynamic environment where external data evolves continuously. As data sources change, versioning is required to capture differences between updates of retrieval resources. 
Input and output data of each component, primarily the retriever (e.g., linking user queries to retrieved documents, embedding IDs, and any filters or reranking applied) and generator (e.g., reconstructing the prompt for generation, including the user query and retrieved context), along with their structures, should also be thoroughly documented to maintain transparency and reproducibility across experiments and production workflows. 

Such traceability information must be securely recorded~\cite{liu2024blockchain} to facilitate debugging and to identify issues related to specific data versions or retrieval source updates. This approach ensures the reliability and accountability of the RAG system over time.

\textbf{Reliability.} The reliability of RAG systems can be affected by unpredictable factors such as bandwidth usage and operational costs. For instance, when users upload large documents, it can impact system performance. 
RAGOps is required to effective manage such operational challenges to ensure reliability of RAG system. Operational strategies need to be implemented, such as load balancing, auto scaling, and prioritization mechanisms to ensure consistent performance even under varying workloads.

\subsection{Human-in-the-Retrieval}

Incorporating human-in-the-loop in the RAG process necessitates redesigning and extending 
the RAG system to effectively integrate valuable but fragmented human feedback. From an operational perspective, it is also crucial to support the continuous incorporation of accumulated human feedback into passive retrieval sources, which influences reasoning and decision-making processes. Tracing updates in human feedback and assessing their impact on 
the RAG system introduces additional complexity. 

\textbf{Adaptability.} There are two key design considerations for storing human feedback as a passive retrieval source for adaptation at runtime. The first involves deciding whether to integrate fragmented tacit knowledge into existing domain knowledge or to store it separately. Human-contributed intermediate knowledge can be merged into existing sources, such as defining the semantic meaning of specific data columns in a tabular structure or adding missing relationships in a knowledge graph. However, if the feedback contains confidential or sensitive information—such as unpublished ideas in scientific contexts—it may be more appropriate to store it separately as a distinct retrieval source. The second consideration concerns the format or data structure used to store the feedback, particularly when it is kept separate. Depending on the nature and volume of the feedback, it may be embedded in a vector database or organized within a knowledge graph.

\textbf{Observability.} It is crucial to consistently monitor human feedback and evaluate its impact on the ongoing RAG process. These assessments are essential for maintaining the overall quality of RAG, particularly when tacit knowledge from human experts has not yet been integrated into existing passive retrieval sources. To enhance real-world applicability and reliability, the testing and evaluation capabilities of RAGOps must be extended with a comprehensive suite of test datasets and the incorporation of user feedback into evaluation cycles.

\textbf{Traceability.} Traceability includes tracking human contributions as inputs to each component, as well as the outputs produced by each component. This includes the user’s input prompt, intermediate outputs generated for querying retrieval sources, feedback provided by human experts, and the generator's final output.

\section{Overarching RAGOps Challenges}
\label{sec_challenges}

\subsection{Responsible and Safe AI}

As with any LLM compound system, RAG applications must adhere to responsible AI guidelines and legal frameworks to ensure responsible AI use~\cite{lu2024responsible}, and the same applies to RAGOps. This adherence involves several critical levels:

First is regulatory compliance. RAGOps must align with regulations such as GDPR and the EU AI Act, along with their codes of practice. These codes articulate commitments, measures, and key performance indicators (KPIs) for stakeholders across the AI supply chain. These measures are applicable and extendable to RAGOps. For instance, the transparency information required by regulations can be automated through RAGOps, streamlining compliance processes. 

Second is alignment with established standards. Compliance with AI safety standards\footnote{``The 10 guardrails,'' Voluntary AI Safety Standard, Australian Government Department of Industry, Science and Resources, accessed 5 April 2025, \url{https://www.industry.gov.au/publications/voluntary-ai-safety-standard/10-guardrails}} is also essential. Certain voluntary guardrails directly influence RAGOps design, such as: \textit{``Test AI models and systems to evaluate model performance and monitor the system once deployed.''}, and \textit{``Keep and maintain records to allow third parties to assess compliance with guardrails.''}. 

Third is adherence to responsible AI principles. RAG applications should incorporate high-level responsible AI principles\footnote{``Australia's AI Ethics Principles,'' Australian Government Department of Industry, Science and Resources, accessed 5 April 2025,  \url{https://www.industry.gov.au/publications/australias-artificial-intelligence-ethics-principles/australias-ai-ethics-principles}}, including fairness, accountability, transparency, and inclusivity. From the RAGOps perspective, emphasis shifts toward the automated testing, evaluation, and monitoring of these principles, ensuring they are upheld throughout the system's lifecycle.

To implement responsible AI principles and standards, it is necessary to build responsible/safe-AI-by-design~\cite{lu2023responsible} into the RAG applications through design patterns and tactics. For example, a black box recorder can be designed to collect critical data of the RAG process at runtime for observability.

\subsection{Lack of Standard Evaluation Metrics}

The absence of standardized evaluation metrics across various functionalities, including retrieval resources, retrieval processes, and generation, creates significant challenges in accurately assessing each of the components and the RAG application as a whole~\cite{gao2024retrievalaugmentedgenerationlargelanguage}. This lack of uniformity hinders the ability to gain a comprehensive understanding of the system’s overall performance. While various prompting techniques have been developed to enhance RAG performance, identifying the most effective approach for specific retrieval sources remains a complex and unresolved challenge.

The design decisions made during the early stages, such as chunking strategies and the choice of embedding models, influence subsequent stages of the data lifecycle and RAGOps. Defining metrics that can effectively assess the impact of these design choices presents a considerable challenge. 

\subsection{Continuous Improvement Through Observability}

As RAGOps systems evolve continuously, observability is required to monitor, assess, and enhance their operations. Observability provides insights into system performance, data quality, and user interactions, which indicates iterative updates and refinements. Decoupling and articulating various aspects of monitorability, including data quality, model quality, and software quality, is challenging. It is crucial to determine the most appropriate monitoring techniques for each aspect to ensure comprehensive oversight of the RAG application. While existing techniques address monitoring for individual RAG components, ensuring seamless monitoring of the entire RAG pipeline and its interactions with the RAG system is critical. Monitoring changes in data sources and understanding their impact on the RAG system add further complexity.

Key challenges in observability include identifying components responsible for detecting performance degradation and designing their interfaces within the system. An important aspect of RAG system is the need to continuously integrate user input and feedback into retrieval process and future performance improvement. This feedback loop can significantly influence observability requirements for individual RAG components, potentially driving new architectural decisions in future iterations. Moreover, extending existing monitoring tactics beyond typical concerns like performance and resource consumption is essential to address the unique characteristics of RAG system, like relevance, groundedness and factuality. 

\subsection{Multimodality}

In real-world application scenarios, such as healthcare, data are inherently multimodal with various types, scales, and formats. For instance, medical records often combine structured tabular data, imaging, and unstructured text, each providing unique insights into patient care. While multimodality is considered a critical element of intelligence and a cornerstone for achieving comprehensive understanding, RAG applications and frameworks with multimodal capabilities are still emerging. Most existing solutions operate with fixed source and target modalities, such as text-to-image, image-to-text, or image-text pair to image retrieval. Some approaches~\cite{lim2024unirag, yu2023unified, liu2023mmhqa} attempt to unify multimodal sources into text, leveraging language models to generate responses. However, this conversion process may lead to the loss of critical information.

Multimodal RAG systems can outperform single-modality models by integrating and analyzing information from various sources. This integration enhances tasks such as anomaly detection, where combining modalities can lead to richer and more robust insights than relying on a single data type.

\subsection{Human-in-the-Retrieval}

A major challenge in integrating humans into the retrieval process lies in determining when their input is necessary~\cite{xu2025humanretrieval}. The aim is to involve humans selectively and strategically, particularly in cases involving critical information, edge cases, closely ranked outputs, or high uncertainty in AI-driven steps. This adds a layer of necessity assessment to the RAG workflow, effectively positioning human expertise as a dynamic and adaptive retrieval resource. Additionally, tools for optimizing human inputs may be required to refine their feedback, converting it into meaningful intermediate contributions that enhance the retrieval process.

Another  challenge is extending human-in-the-loop functionality to other components of the RAG application beyond retrieval, such as orchestration and generation. This becomes particularly relevant in high-uncertainty scenarios where expert intervention is needed to assess workflow quality, evaluate retrieved contexts, and validate query formulations. Human corrections and adjustments based on these evaluations play a crucial role in improving the overall effectiveness and accuracy of both the retrieval and generation stages.

\section{Use Cases}
\label{casestudies}

\subsection{Taxation Assistant}

We collaborated on a project with a local AI startup, which recently launched an LLM-based taxation assistant aimed at helping tax professionals analyze complex tax scenarios. The application provides instant answers with explanations, integrates real-time updates from government service departments, and performs automatic reference checks. As part of the project, we assessed their system design and enhanced their application by incorporating an RAGOps infrastructure.

\subsubsection{RAG Design and Development}

The initial design decisions on \textit{retrieval sources} mainly involved selecting the approach for incorporating domain-specific taxation data, specifically the open data provided by the government department, including \textit{topics} and \textit{types}. \textit{Topics} serve as a fixed data source, while \textit{types} are organized according to the taxation/law hierarchy, which are updated on a weekly basis. A web crawler is used to collect data from the government website, with crawling scheduled weekly. Leveraging authorized open data from government significantly reduces the likelihood of hallucinations. Using RAG to dynamically query the data ensures that any link to the government data can easily be removed by the company if needed. 

The company made a decision between using embedding or knowledge graph to capture the authoritative data from government. To construct the retrieval source, plain text is stored in a vector database, as the primary functionality is to identify relevant information. In addition, the focus is specifically on taxation regulations without complex dependencies on other financial legislation, thus, using a knowledge graph was unnecessarily heavy and not cost-efficient for this case.

When determining granularity and metadata, the company chose to define each question and answer under a private ruling (found on a single web page) as a data chunk, which is then constructed as an embedding. The size of these chunks was carefully considered, balancing the trade-off between cost and efficiency --- for instance, whether a chunk should represent an entire web page or just a section of it. Metadata selection was primarily driven by the scenarios in which tax professionals would use the assistant. For specific taxation cases, metadata includes the decision and the reasoning behind it. For most other documents, the metadata consists of a summary of the document. 

For the \textit{retriever}, The company implemented a classifier to categorize users' queries into different types of scenarios, and utilizes scenario-specific prompt templates to help the LLM better understand and interpret queries. A query generator leverages the output of these templates to transform the query into a vector representation, which is then compared against vectors of stored chunks in the vector database. This process retrieves two types of information: \textit{rulings} and \textit{legislation}.

As the retriever provides an initial list of ranked results to the generator, the company introduced a reranker to further assess their relevance. This reranker employs an LLM-based evaluator to refine the rankings. A privacy guardrail was also built within the retriever to desensitize sensitive user data before processing queries, ensuring privacy and security throughout the retrieval process.

The \textit{generator} has a context constructor that utilizes response templates to structure recommendations, enhancing readability and helping tax professionals understand the reasoning behind the generated responses. Another key decision was to incorporate user profile, which can be created by having users input their information or by continuously learning and updating the profile based on interactions. The generator uses the information to refine the output, ensuring personalized and relevant results.  

\subsubsection{RAGOps}

During the design phase of the taxation assistant, three quality attributes were identified that are partially addressed and considered from operational perspectives:

\begin{itemize}
    \item Adaptability: Incorporating user feedback into the assistant to improve future recommendation generation. 
    \item[] \textit{User story}: A tax professional or a runtime evaluator identifies and submits feedback pointing out a mistake in a recommendation on capital gains tax. The tax assistant integrates the feedback into its agent memory and adapts its reasoning logic for similar future scenarios involving capital gains tax.
    \item Observability: Enabling stakeholders to track historical queries, monitor user feedback, and receive alerts on assistant health, performance or users complaints. \item[] \textit{User story}: An alert is triggered due to an increasing number of low scores provided by the assistant user about recommendation on small business tax deductions. The tax assistant automatically logs all queries, responses, and user feedback.
    \item Contestability: Allowing users to challenge the assistant's recommendation and submit feedback for review. 
    \item[] \textit{User story}: A tax professional disagrees with the recommendation provided on superannuation contributions and submits feedback to challenge the recommendation. The tax assistant logs the feedback and flags the case for review by tax experts.
\end{itemize}

To support and fulfill these quality requirements, we introduce an RAGOps infrastructure layer that cross-cuts all RAG components, including retrieval sources, retriever and generator. This layer logs all queries, responses, user feedback, and the input and output of each component, providing comprehensive visibility into the assistant's operations and improving monitoring and evaluation capabilities. 

\subsubsection{Data management}

For testing and evaluation, we compiled a set of ground truth data from questions and corresponding certified replies available on the national tax office’s open forum. For complex scenarios, we initially gathered ground truth data directly from the company’s domain experts. Subsequently, we used the LLM to generate additional ground truth data for evaluation purposes.

\subsection{Magda Copilot}

Magda\footnote{``Magda data catalog
,'' CSIRO, accessed 5 April 2025, \url{https://www.csiro.au/en/research/technology-space/cyber/Magda-data-catalog}} is a data catalog system that serves as a centralized hub for cataloging an organization’s data. It provides open-source software to assist organizations or government agencies in managing data tasks such as collection, authoring, discovery, usage, sharing across organizations, or publishing to open data portals.

In scientific contexts, our internal scientists from various domains utilize the open data available on the platform. Within a multidisciplinary project, we collaborate closely with scientists in agriculture and biology to explore and understand their requirements for using LLMs with domain-specific data sources via RAG to address their daily scientific questions. From these observations and insights, we developed the Magda Copilot, a tool designed to help scientists explore and discover data on the Magda platform.

\subsubsection{RAG Design and Development}

The Magda Copilot is designed to autonomously integrate with scientific tools commonly used by researchers. The first stage incorporates five primary tools:

\begin{itemize}
    \item Basic info tool: the tool retrieves basic information about proteins from UniProt database\footnote{UniProt, accessed 5 April 2025, \url{https://www.uniprot.org}}.
    \item Sequence similarity tool: the tool retrieves proteins with similar sequences using a BLAST (Basic Local Alignment Search Tool)\footnote{``Basic Local Alignment Search Tool'', National Center for Biotechnology Information, accessed 5 April 2025, \url{https://blast.ncbi.nlm.nih.gov/Blast.cgi}} search based on a UniProt accession code.
    \item Biochemical characteristics tool: the tool can generate SQL queries to retrieve information from an internal tabular dataset about the biochemical characteristics of proteins. 
    \item Structure tool: the tool utilizes AlphaFold Protein Structure Database to retrieve and display structural information about proteins based on UniProt accession code.
    \item Causal graph discovery tool: the tool extracts causal relationships from scientific literature stored as a repository of pdf files.
\end{itemize}

A sample and typical scenario involves identifying enzymes that play a role in breaking down polyethylene terephthalate (PET). For example, the process begins with a known enzyme with UniProt code A0A0K8P6T7, which is  already recognized for PET degradation. Its basic information is retrieved using its UniProt code. Next, the sequence similarity tool is used to find enzymes having similar sequences. The basic information and the biochemical characteristics are then queried to check whether each candidate enzyme is active on PET under certain conditions. As such information can be incomplete in the data, the structure information of these enzymes are used to further identify whether active sites exist that are likely to degrade PET. These capabilities are further enhanced by the causal graph discovery tool which visualize a causal graph of potential factors involved in the PET degradation activities to complement insights gained from sequence and structural analysis. The Magda Copilot is able to autonomously execute the workflow, and after identifying several candidates, they are ultimately validated through experimental studies conducted by researchers. 

The \textit{retriever} in the system routes user queries to corresponding tools by identifying and decomposing user's query into atomic queries.
It generates prompts tailored to the five tools mentioned. The \textit{generator} then synthesizes the final output based on results returned by these tools, providing customization applied to handle domain-specific terminology and contextual accuracy. Additionally, Magda Copilot and its tools are wrapped as a suite of REST API services, enabling straightforward integration with other platforms.

\subsubsection{RAGOps}

Through the multidisciplinary project, we observed that when domain experts use RAG systems, their questions often lack sufficient context. Furthermore, domain experts frequently possess tacit knowledge --- implicit understanding that is challenging to articulate or extract. This tacit knowledge cannot easily be captured by passive retrieval sources through written or verbal communication, However, it can be integrated as external data to enhance retrieval quality or guide the workflow executed by Magda Copilot. Consequently, interactivity emerges as a key quality attribute. In addition, we identified three other critical quality attributes that need to be addressed from an operational perspective:

\begin{itemize}
    \item Interactivity: The application should facilitate effective user interaction, enabling the Magda Copilot to engage end users for further information or feedback when needed.
    \item Observability: When user queries are processed through a series of tools, potentially incorporating tacit knowledge, each step impacts the final result and the overall performance of the retrieval process. Monitoring and observing the performance of each step and the end-to-end process is essential.
    \item Traceability: In scientific discovery scenarios, the ability to trace the query history is crucial. This includes logging the tools selected, as well as the input and output of each step. Comprehensive logging of all the actions ensures transparency and reproducibility. 
    \item Adaptability: Gradually incorporating tacit knowledge collected during interactions into the passive retrieval sources is important for improving retrieval quality over time. Moreover, more data sources or tools may be further integrated into the system, requiring the retrieval process to adapt accordingly.
\end{itemize}

\subsubsection{Data management}

To meet these quality requirements, the operational design of Magda Copilot incorporates specific functionalities. The structure tool retrieves the 3D structure information of proteins from AlphaFold protein structure database. When the proteins of interests are not included in the database, the system will run the structure prediction using AlphaFold and store the results locally for future use. By leveraging both online and local databases, the structure tool ensures seamless integration and accessibility of new structure data as it becomes available. This design enables efficient and comprehensive capture of results, enhancing the overall utility and adaptability.

Other data sources are continuously expanded to enhance the capabilities of Magda Copilot, and the data within each source is also continuously enriched to incorporate the latest scientific research progress. This may introduce new forms of data, and the retrieval process must adapt to accommodate these changes effectively. The performance of the Magda Copilot is assessed using a combination of quantitative and qualitative methods. Feedback from domain scientists play a critical role in validating the relevance and accuracy of retrieved and generated outputs. Additionally, user satisfaction from the domain scientists capture provide valuable insights into usability and effectiveness from the end user's perspective.

\noindent{\textit{Tacit Knowledge Management}}: Domain scientists play a key role in refining retrieval and generation outputs by providing invaluable feedback and contributing to iterative improvement processes. Their expertise helps uncover tacit knowledge that is not captured through traditional documentation. Tacit knowledge can be extracted from the interaction with domain experts, such as the correct workflow for analyzing specific data, the effective way for conducting data search, and the reasoning behind these approaches. This knowledge gradually evolves into a new data source that guides Magda Copilot in executing research workflows. Given the varied granularity and diverse perspectives of this knowledge, iterative distillation is necessary to refine and structure it into a usable form. By leveraging expert knowledge and automating complex processes automatically, Magda Copilot has assisted the discovery of new research insights.

\section{Summary and Future Work}
\label{sec_summary}

Retrieval-Augmented Generation (RAG) has emerged as a promising solution to address key challenges faced by LLMs (Large Language Models), such as hallucination, outdated or non-removable parametric knowledge, and non-traceable reasoning processes. Current LLMOps tools predominantly focus on model management, offering limited support for data-related aspects, particularly the data retrieved after an LLM is deployed. 

This paper conceptualizes RAGOps, which builds upon LLMOps by emphasizing robust data management to address the dynamic nature of external data sources. This extension necessitates automated evaluation and testing methods to improve data operations, ensuring enhanced retrieval relevance and generation quality. This paper characterizes the generic architecture of RAG applications using the 4+1 model view of software architectures, outlines the integrated lifecycle of query processing pipeline by combining LLM and data management lifecycles, defines the key design considerations and corresponding quaility tradeoffs of RAGOps across different stages, identifies research challenges associated with each stage, and presents two practical use cases of RAG applications, offering valuable insights for advancing the operationalization of RAG applications.

Several directions remain open for future work. The verification phase (Section~\ref{subsubsec:verification}) of the knowledge base maintenance lacks a mechanism to detect the malicious insertion of detrimental data or other forms of attacks. Specific semantic checks complementing classical cybersecurity measures constitute an avenue for future research. The applicability of recent developments in agentic approaches, which could equip RAG systems with extensive autonomy, also merits further investigation. Another direction is the development of tooling support tailored to RAGOps. Dedicated tools are needed to assist developers in monitoring data retrieval behaviors, tracing generated outputs to their sources, and evaluating the performance impact of data changes. Such tooling can significantly lower the barrier to adoption and foster best practices in managing the full lifecycle of RAG systems.

\bibliographystyle{elsarticle-num}
\bibliography{references}  

\end{document}